# Crosstalk Noise Modeling for RC and RLC interconnects in Deep Submicron VLSI Circuits

P.V.Hunagund, A.B.Kalpana

**Abstract**—The crosstalk noise model for noise constrained interconnects optimization is presented for RC interconnects. The proposed model has simple closed-form expressions, which is capable of predicting the noise amplitude and the noise pulse width of an RC interconnect as well as coupling locations (near-driver and near-receiver) on victim net. This paper also presents a crosstalk noise model for both identical and non identical coupled resistance–inductance–capacitance (RLC) interconnects, which is developed based on a decoupling technique exhibiting an average error of 6.8% as compared to SPICE. The crosstalk noise model, together with a proposed concept of effective mutual inductance, is applied to evaluate the effectiveness of the shielding technique.

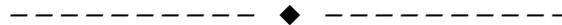

## 1 INTRODUCTION

THE minimum feature size in circuits is shrinking, signal integrity issues gain importance due to increased coupling between nets in VLSI circuits, this coupling that may result in cross-talk noise. Decreasing feature size affects the crosstalk noise problem and also affects on design's timing and functionality goals [1] [2].

This paper proposes a much improved crosstalk noise model, called the 2π- *model* taking into consideration many key parameters, such as the aggressor slew at the coupling location, the coupling location at the victim net (near-driver or near-receiver), and the coarse distributed RC characteristics for victim net. The 2π model is very accurate, with less than 6% error on average compared with HSPICE simulations.

With faster rise times and lower resistance, long wide wires in the upper metal layers exhibit significant inductive effects. An efficient resistance–inductance–capacitance (*RLC*) model of the on-chip interconnect is therefore critical in high-level design, logic synthesis, and physical design. A closed-form expression for the crosstalk noise between two identical *RLC* lines is developed in [3], assuming that the two interconnects are loosely coupled. In [4], a technique to decouple coupled *RLC* interconnects into independent interconnects is developed based on a modal analysis. This decoupling method, however, assumes a TEM mode approximation, which is only valid in a two-dimensional structure with a perfect current return path in the ground plane directly beneath the conductors [5]. An estimate of crosstalk noise among multiple *RLC* interconnects is required to efficiently implement shielding techniques. Inserting shield lines can greatly reduce both capacitive coupling [6] and mutual inductive coupling by providing a closer current return path for both the aggressor and victim lines. An efficient estimate of the crosstalk noise between coupled interconnects including the effect of shield insertion is therefore critical during the routing and verification phase to guarantee signal integrity.

## 2 CROSSTALK NOISE MODELING FOR RC INTERCONNECTS USING 2-π MODEL

### 2.1 2-π Model and its Analytical Waveform

For simplicity, we first explain 2-π model for the case where the victim net is an RC line. For a victim net with some aggressor nearby, as shown in Fig. 1 (a), let the aggressor voltage pulse at the coupling location be a saturated ramp input with transition time (i.e., slew) being $t_r$ and the interconnect length of the victim net before the coupling, at the coupling and after the coupling be $L_s$, $L_c$ and $L_e$, respectively. The 2-π type reduced RC model is generated as shown in Fig.1 (b) to compute the crosstalk noise at the receiver. It is called 2-π model because the victim net is modeled as two π-type RC circuits, one before the coupling and one after the coupling. The victim driver is modeled by effective resistance $R_d$, other RC parameters $C_x$, $C_l$, $R_s$, $C_2$, $R_e$, and $C_L$ are computed from the geometric information from Fig.1 (a) in the following manner.

- P.V.Hunagund is with the Department of Applied Electronics, Gulbarga University, Gulbarga, INDIA.
- A.B.Kalpana is with the Department of Electronics and Communication, Bangalore Institute of Technology, Bangalore, INDIA.

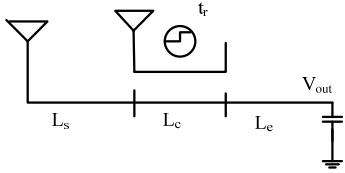

Fig 1(a) Layout of a victim net and aggressor above it.

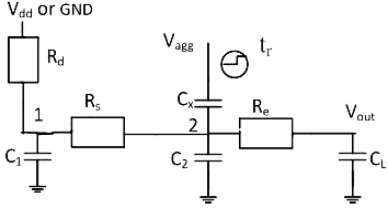

Fig. 1(b) 2π crosstalk noise model.

The coupling node (node 2) is set to be the center of the coupling portion of the victim net, i.e., $L_s + L_c/2$ from the source. Let the upstream and downstream interconnect resistance capacitance at Node 2 be $R_s/C_s$ and $R_e/C_e$, respectively. Then capacitance values are set to be $C_1 = C_s/2$, $C_2 = (C_s + C_e)/2$ and $C_L = C_e/2 + C_1$. Compared with [7,8] which only used one lumped RC for the victim net, it is obvious that our 2-π model can model the *coarse* distributed RC characteristics. In addition, since we consider only those *key* parameters, the resulting 2-π model can be solved analytically.

From Fig.1(b), we have the impedance at node 1, $Z_1$ Satisfying the following

$$\frac{1}{Z_1} = \frac{1}{R_d} + sC_1$$

Then at node 2, we have

$$\frac{1}{Z_2} = \frac{1}{(Z_1 + R_S)} + sC_2 + \frac{1}{R_e + \frac{1}{sC_L}}$$

Denote the s-domain voltage at node 2 by $V_2(s)$, then

$$V_2(s) = \frac{Z_2}{Z_2 + \frac{1}{sC_L}} \cdot V_{agg}(s)$$

The output voltage $V_{out}$ in the s-domain is

$$V_{out}(s) = V_2(s) \cdot \frac{\frac{1}{sC_L}}{R_e + \frac{1}{sC_L}} \quad (1)$$

Substituting $Z_1$, $Z_2$ and $V_2$ into $V_{out}(s)$, we have

$$V_{out}(s) = \frac{a_2 s^2 + a_1 s}{s^3 + b_2 s^2 + b_1 s + b_0} \cdot V_{agg}(s) \quad (2)$$

Where the coefficient are

$$a_2 = \frac{K_1}{K_2}, \quad a_1 = \frac{(R_d + R_s)C_x}{K_2}$$

$$b_2 = \frac{((C_2 + C_x)(R_e C_L(R_d + R_S) + R_d R_s C_1) + R_d R_e C_1 C_L + C_L R_d R_s C_1)}{K_2}$$

$$b_1 = \frac{((R_d + R_s)(C_x + C_2 + C_L) + (R_e C_L + R_d C_1))}{K_2}$$

$$b_0 = \frac{1}{K_2}, \quad K_1 = C_x R_d R_s C_1$$

$$K_2 = R_d R_s C_1 C_L R_e (C_x + C_2)$$

Writing the transfer function H(s) into the poles/residues form:

$$H(s) = \frac{a_2 s^2 + a_1 s}{s^3 + b_2 s^2 + b_1 s + b_0} \equiv \frac{k_1}{s - s_1} + \frac{k_2}{s - s_2} + \frac{k_3}{s - s_3}$$

The three poles $s_1, s_2$ and $s_3$ are the three roots of $s^3 + b_2 s^2 + b_1 s + b = 0$, which can be obtained analytically using standard mathematical techniques. After each poles/residue pair is obtained, its corresponding time domain function is just $f_i(t) = k_i e^{s_i t}$ (i = 1, 2, 3).

For the aggressor with saturated ramp input with normalized $V_{dd} = 1$ and transition time $t_r$, i.e.

$$v_{agg} = \begin{cases} t/t_r & 0 \le t \le t_r \\ 1 & t \ge t_r \end{cases}$$

Its Laplace transformation is

$$V_{agg}(s) = \frac{1 - e^{-st_r}}{s^2 t_r} \quad (3)$$

Then for each pole/residue pair, the s-domain output $V_{out}(s) = \frac{k_i}{s - s_i} V_{agg}(s)$ and its inverse Laplace is just the convolution of $f_i(t)$ and $g(t)$

$$V_{out}(t) = f_i(t) * g(t) = \int_0^t f_i(t-u) g(u) du,$$

$$= \begin{cases} -\frac{k_i(1 + s_i t)}{s_i^2 t_r} + \frac{k_i e^{s_i t}}{s_i^2 t_r} & 0 \le t \le t_r \\ -\frac{k_i e^{s_i(t - t_r)}}{s_i^2 t_r} + \frac{k_i e^{s_i t}}{s_i^2 t_r} + \frac{k_i}{s_i} & t \ge t_r \end{cases} \quad (4)$$

Therefore, the final noise voltage waveform is simply the summation of the voltage waveform from each pole/residue pair

$$v_{out}(t) = v_{out1}(t) + v_{out2}(t) + v_{out3}(t) \quad (5)$$

The 2-π model has been tested extensively and its waveform from (5) can be shown to be almost identical compared to HSPICE simulations.

## 2.2 Closed-Form Noise Amplitude and Width

In this subsection, we will further simplify the original *2-π* model and derive closed-form formulae for noise amplitude and noise width.

Using dominant-pole approximation method in a similar manner like [9, 10, 11], we can simplify (2) into

$$V_{out}(s) \approx \frac{a_1 s}{b_1 s + b_0} \cdot V_{agg}(s) = \frac{t_x \left(1 - e^{-st_r}\right)}{st_r(st_v + 1)} \quad (6)$$

Where the coefficient are

$$t_x = (R_d + R_s)C_x \quad (7)$$

$$t_v = (R_d + R_s)(C_x + C_2 + C_L) + (R_e C_L + R_d C_1) \quad (8)$$

It is interesting to observe that $t_x$ is in fact the RC delay term from the upstream resistance of the coupling element times the coupling capacitance, while $t_v$ is the distributed Elmore delay of victim net. We will further discuss their implications later on.

Computing the inverse Laplace transform of (6), we can obtain the simple time domain waveform

$$v_{out} = \begin{cases} \dfrac{t_x}{t_r}\left(1 - e^{-t/t_v}\right) & 0 \le t \le t_r \\[6pt] \dfrac{t_x}{t_r}\left(e^{-(t-t_r)/t_v} - e^{-t/t_v}\right) & t > t_r \end{cases} \quad (9)$$

It is easy to verify that in the above noise expression, $v_{out}$ monotonically increases at $0 \le t \le t_r$, and monotonically decreases at $t > t_r$. So the peak noise will be at $t = t_r$, with the value of

$$v_{max} = \dfrac{t_x}{t_r}\left(1 - e^{-t_r/t_v}\right) \quad (10)$$

It is also interesting to compare with the recent work by [12], where the peak noise with saturated ramp input can be written as $v'_{max} = t_x/(t_v + t_r/2)$. Although obtained from a totally different approach, $v'_{max}$ from [12] is indeed a first-order approximation of our $v_{max}$ in (10), since. However, such approximation is only when $t_r < t_v$. It will be much off when $t_r \gg t_v$. This explains why $v'_{max}$ in [12] gives twice peak noise.

$$\dfrac{t_x}{t_r}\left(1 - e^{-t_r/t_v}\right) = \dfrac{t_x}{t_v}\left[1 - \dfrac{1}{2}\dfrac{t_r}{t_v} + \ldots\right] \quad (11)$$

$$\dfrac{t_x}{t_v}\dfrac{1}{1 + \dfrac{1}{2}\dfrac{t_r}{t_v}} = \dfrac{t_x}{t_v + \dfrac{t_r}{2}} \quad (12)$$

Peak noise amplitude $v_{max}$ is not the only metric to characterize noise. Under some circumstance, even the peak noise exceeds certain threshold voltage; a receiver may still be noise immune. This can be characterized by some noise amplitude versus width plots.

*Noise Width:* Given certain threshold voltage level $v_t$, the noise width for a noise pulse is defined to be the length of time interval that noise spike voltage $v$ is larger or equal to $v_t$.

From (9), we can compute $t_1$ and $t_2$, and thus the noise width

$$t_2 - t_1 = t_v \ln\left[\dfrac{(t_x - t_r v_t)(e^{t_r/t_v} - 1)}{t_r v_t}\right] \quad (13)$$

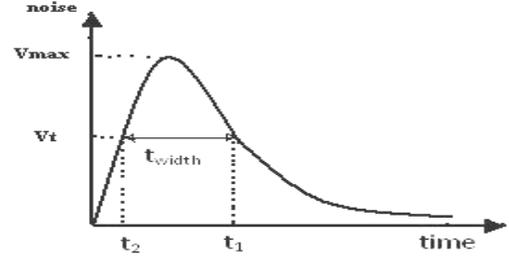
Fig.2: Illustration of the noise width.

In this paper, we set the threshold voltage $v_t$ to be half of the peak noise voltage, $v_t = v_{max}/2$. Then the noise width of (13) is simplified into

$$t_{width} = t_2 - t_1 = t_r + t_v \ln\left[\dfrac{1 - e^{-2t_r/t_v}}{1 - e^{-t_r/t_v}}\right] \quad (14)$$

## 3 CROSSTALK NOISE MODELING FOR RLC INTERCONNECTS USING DECOUPLING TECHNIQUE

Two-coupled *RLC* interconnects with a coupled capacitance per unit length $c_c$, mutual inductance $l_m$, resistance $r(1+\Delta r)$ and $r(1-\Delta r)$, self-inductance $l(1 + \Delta l)$ and $l(1 - \Delta l)$, and ground capacitances $c_g(1 + \Delta c)$ and $c_g(1 - \Delta c)$, respectively, are shown in Fig. 3.

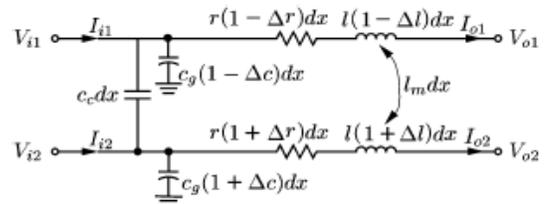
Fig.3 Infinitesimally small segment of two coupled *RLC* interconnects.

The *ABCD* matrix E, for an infinitesimally small segment of these two coupled interconnects can be obtained, as shown in (15). Furthermore, the matrix E can be diagonalized as

$$E = \begin{bmatrix} 1 & 0 & [r(1-\Delta r)+s l(1-\Delta l)]dx & s l_m dx \\ 0 & 1 & s l_m dx & [r(1+\Delta r)+s l(1+\Delta l)]dx \\ s[c_g(1-\Delta c)+c_c]dx & -s c_c dx & 1 & 0 \\ -s c_c dx & s[c_g(1+\Delta c)+c_c]dx & 0 & 1 \end{bmatrix} \quad (15)$$

$$E = W \Lambda W^{-1} \quad (16)$$

Where

$$\Lambda = \begin{bmatrix} (1-\theta_1 dx) & 0 & 0 & 0 \\ 0 & (1+\theta_1 dx) & 0 & 0 \\ 0 & 0 & (1-\theta_2 dx) & 0 \\ 0 & 0 & 0 & (1+\theta_2 dx) \end{bmatrix} \quad (17)$$

$$W = \begin{bmatrix} -Z_{01} & Z_{01} & Z_{02} & -Z_{02} \\ -Z_{01} & Z_{01} & -Z_{02} & Z_{02} \\ 1 & 1 & -1 & -1 \\ 1 & 1 & 1 & 1 \end{bmatrix} \quad (18)$$

In general, $\theta_1$ and $\theta_2$ are functions of the interconnect impedance parameters (resistances, capacitance, and inductances) and are difficult to solve analytically. If non identical coupled interconnects are part of a bus structure with the same width, height, and length, the resistance of the two non identical interconnects are equal, i.e., $\Delta r = 0$. Under the condition of $\Delta r = 0$, and a moment matching approximation $\theta_1$, $\theta_2$, $Z_{o1}$, and $Z_{o2}$ can be approximated as

$$\theta_1 = \sqrt{sC'_g(r + s(l' + l'_m))} \quad (19)$$

$$\theta_2 = \sqrt{s(C'_g + 2C'_c)(r + s(l' - l'_m))} \quad (20)$$

$$Z_{01} = \sqrt{\frac{(r + s(l' + l'_m))}{sC'_g}} \quad (21)$$

$$Z_{02} = \sqrt{\frac{(r + s(l' - l'_m))}{s(C'_g + 2C'_c)}} \quad (22)$$

Where $c_g'$, $c_c'$, $l'$, $l'_m$ are

$$c'_g = c_g \left(1 + \frac{c_c}{c_g} - \sqrt{\frac{c_c^2}{c_g^2} + \Delta c^2}\right) \quad (23)$$

$$c'_c = c_c \sqrt{1 + \frac{c_c^2}{c_g^2} \Delta c^2} \quad (24)$$

$$l' = l \quad (25)$$

$$l'_m = l_m \frac{c_c}{\sqrt{c_c^2 + c_g^2 \Delta c^2}} - l \frac{c_g \Delta c \Delta l}{\sqrt{c_c^2 + c_g^2 \Delta c^2}} \quad (26)$$

The physical meaning of $\theta_2$ ($Z_{o2}$) is the propagation constant (characteristic impedance) of coupled interconnects when both inputs switch in opposite directions. These two decoupled interconnects can therefore be used to determine the output waveforms of two coupled interconnects.

## 4 CROSSTALK NOISE MODEL FOR TWO-COUPLED INTERCONNECTS

Based on the decoupling technique, the crosstalk noise model is first developed for two identical coupled *RLC* interconnects. The crosstalk noise model is then applied to non identical coupled *RLC* interconnects and compared with SPICE, exhibiting an average error of 6.8%.

### 4.1 Crosstalk Noise Model of Two Identical Coupled Interconnects

For the coupled interconnects shown in Fig. 3 with $\Delta r = 0$, $\Delta c = 0$, and $\Delta l = 0$, the transient response at the two outputs can be expressed using the normalized variables listed in Table I. Furthermore, in order to characterize the effect of inductance on the crosstalk noise, a parameter $\zeta$, described in [13], is used, where $\zeta$ is defined as

$$\zeta = \frac{R_T + R_T C_T + R_R C_T + 0.5 R_R}{2\sqrt{(1 + C_T)}} \quad (27)$$

TABLE I. NORMALIZED VARIABLES FOR TWO COUPLED INTERCONNECTS

| Variable | Definition | Physical Meaning |
|---|---|---|
| $Z_o$ | $\sqrt{l/c_g}$ | Characteristic impedance |
| $t_f$ | $h\sqrt{lc_g}$ | Time of flight |
| $R_R$ | $hr/Z_o$ | Normalized line resistance |
| $R_T$ | $R_s/Z_o$ | Normalized driver resistance |
| $C_T$ | $C_L/(hc_g)$ | Normalized load capacitance |
| $K_C$ | $c_c/c_g$ | Normalized coupling capacitance |
| $K_L$ | $l_m/l$ | Normalized coupling inductance |

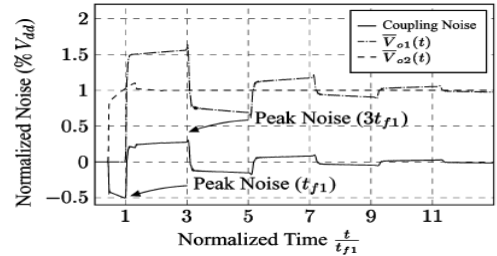

Fig. 4. Output waveform of decoupled interconnects and waveform of coupled noise between two coupled interconnects when t > t (K = 0:769 and K = 0:217).

The input of the victim line remains at ground while the input of the aggressor line is a step input. The crosstalk noise can therefore be expressed using only five variables $\zeta$, $C_T$, $R_T$, $K_C$, and $K_L$. The decoupled interconnects can be used to determine the peak crosstalk noise. For two strongly inductively coupled interconnects ($K_L \gg K_C$ such that $t_{f1} > t_{f2}$), the waveform of the coupling noise and the output waveforms, $V_{o1}(t)$ and $V_{o2}(t)$, of the decoupled interconnects are shown in Fig. 4, where $t_{f1}$ and $t_{f2}$ are

$$t_{f1} = h\sqrt{(l + l_m)c_g} \quad (28)$$

$$t_{f1} = h\sqrt{(l - l_m)(c_g + 2c_c)} \quad (29)$$

$t_{f1}$ and $t_{f2}$ are the times of flight of two decoupled interconnects, respectively.

Based on the traveling-wave model of a transmission line, the traveling wave is reflected at the load, returns to the source, and then returns to the load, causing the output to overshoot and undershoot at the times of $t_f$ and $3t_f$, respectively. During the interval between $t_f$ and $3t_f$, the output of a lossy transmission line with a capacitive load behaves as an RC line, and the output increases due to RC charging [14]. The waveform of the coupling noise can be determined by subtracting the decoupled voltage Vo2(t) from Vo1(t). The negative peak of the coupling noise occurs at time $t_{f1}$, as shown in Fig.4, and is

$$V_{noise}(t_{f1}) = -\frac{1}{2}\overline{V}_{02}(t_{f1}) \quad (30)$$

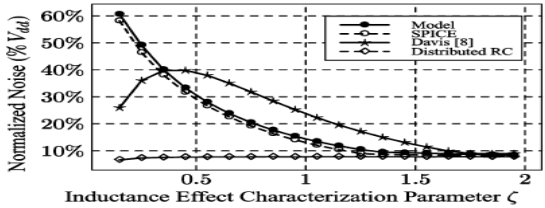

Fig. 5. Comparison of crosstalk model to SPICE, Davis and distributed RC model for different values of ζ (K = 0:217, K = 0:769, C = 0:05, and R = 0:25).

At the time of $3t_{f1}$, the decoupled voltage Vo1(t) is maximum. The positive peak of the coupling noise is

$$V_{noise}(3t_{f1}) = \frac{1}{2}\left(\overline{V}_{01}(3t_{f1}) - \overline{V}_{02}(3t_{f1})\right) \quad (31)$$

Combining (30) and (31), the peak crosstalk noise of two strongly inductively coupled interconnects is

$$V_{peak} = \max\{V_{noise}(t_{f1}), V_{noise}(3t_{f1})\} \quad (32)$$

An analysis of the crosstalk noise when $t_{f1} < t_{f2}$ is similar to an analysis of the crosstalk noise with the positive and negative peak noise occurring at $t_{f2}$ and $3t_{f2}$, respectively. The peak crosstalk noise between two coupled interconnects (either $t_{f1} > t_{f2}$ or $t_{f1} < t_{f2}$) can be unified and is

$$t_{f\max} = \max\{t_{f1}, t_{f2}\} \quad (33)$$

$$V_{peak} = \max\{V_{noise}(t_{f\max}), V_{noise}(3t_{f\max})\} \quad (34)$$

The peak noise in (34) is determined from the transient response of the two decoupled interconnects. In order to determine the precise value of the decoupled voltages Vo1(t) and Vo2(t) at $t_f$ max and $3t_f$ max, a traveling wave-based approximation technique (TWA), as described in [14], is used to construct the transient output response of the two decoupled interconnects. Through the TWA technique, the peak crosstalk noise is compared to SPICE for various values of the five variables ζ, CT, RT, KC, and KL, as shown in Figs. 5 and 6.

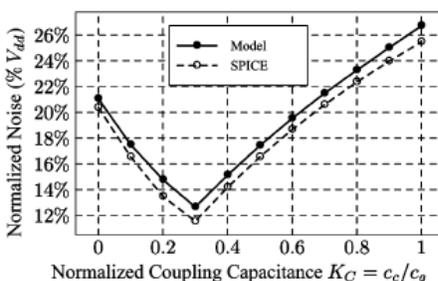

Fig.6. Comparison of crosstalk model to SPICE for different values of K (K = 0:769, ζ = 1, C = 0:05, and R = 0:25).

The interconnect is divided into segments with a length of 10 μm, where each segment is modeled by a π circuit with resistance, ground capacitance, coupling capacitance, partial self-inductance, and partial mutual inductance. The coupled interconnects are simulated using SPICE. The peak crosstalk noise of two coupled RLC interconnects decreases when the inductance effect characterization parameter ζ increases (producing a smaller inductance effect), as shown in Fig. 5. As expected, the distributed RC interconnect model can be used to determine the peak crosstalk noise when ζ is sufficient large (ζ > 1:5). The peak noise is almost constant for the normalized load capacitance CT varying over the practical range of 0 < CT < 0.1, and decreases with larger normalized driver resistance RT. The peak crosstalk noise does not increase monotonically with an increase in the normalized inductive coupling factor $K_L$ or capacitive coupling factor $K_C$ (as shown in Fig. 6).

## 5 CONCLUSION

This paper presents an much improved 2-π crosstalk noise model for RC interconnects with less than 6% error on average compared with HSPICE simulation, for both peak noise voltage and noise width estimations. This paper also assumes a saturated ramp input for the aggressor net and also closed-form expression for peak noise using 2π model under the exponential aggressor input. This 2π model will be useful in many other applications at various levels to guide noise-aware DSM circuit designs.

A crosstalk noise modeling for RLC interconnects using decoupling technique for both identical and non identical coupled RLC interconnects is developed based on the ABCD matrix of interconnects. Based on the decoupling technique, an analytic crosstalk noise model is presented, with the peak noise occurring at the time of flight $t_f$ or $3t_f$. The model exhibits an average error of 6.8% as compared to SPICE.

**P.V.Hunagund** received his post graduation in 1981,Ph.D degree in 1992 from Gulbarga University, Currently he is Chairman Department of Applied Electronics, Gulbarga University, Gulbarga, INDIA, his main research interests include Microwave Electronics, VLSI Design, Microcontrollers, and Instrumentation.

**A.B.Kalpana** received B.E, Degree from Bangalore University, In1995, M.E, Degree from U.V.C.E, Bangalore in 2001, pursuing Ph.D in the Department of Applied Electronics, Gulbarga University, Gulbarga, INDIA, currently she is working as senior grade Lecturer in the Department of Electronics and communication, Bangalore Institute of Technology, Bangalore, INDIA, her research interests include Analysis and Design of VLSI circuits and Power Electronics.